\documentclass[twoside]{LCWS11}
\usepackage[latin1]{inputenc}
\usepackage[dvips]{graphicx,epsfig,color}
\usepackage{wrapfig,rotating}
\usepackage{amssymb,amsmath,array}
\usepackage{cite}
\include{paperdef}
\graphicspath{{figs/}}

\begin{document}

\thispagestyle{empty}
\setcounter{page}{0}
\def\thefootnote{\fnsymbol{footnote}}

\begin{flushright}
\mbox{}
CERN--PH--TH/2012--022 \\
KA--TP--03--2012 
\end{flushright}

\vspace{1cm}

\begin{center}

{\large\sc {\bf MSSM Higgs Bosons from Stop and Chargino Decays}}
\footnote{Talk given by S.H.\ at the {\em LCWS 2011}, 
September 2011, Granada, Spain}

\vspace{1cm}

{\sc 
S.~Heinemeyer$^{1}$%
\footnote{
email: Sven.Heinemeyer@cern.ch}%
, F.~v.d.~Pahlen$^{1}$%
\footnote{email: pahlen@ifca.unican.es}%
, H.~Rzehak$^{2}$%
\footnote{email: heidi.rzehak@cern.ch}%
\footnote{on leave from 
Albert-Ludwigs-Universit\"at Freiburg, Physikalisches Institut, D--79104
Freiburg, Germany 
}%
~and C.~Schappacher$^{3}$%
\footnote{email: cs@particle.uni-karlsruhe.de}%
}

\vspace*{1cm}

{\it
$^1$Instituto de F\'isica de Cantabria (CSIC-UC), 
Santander,  Spain 

\vspace{0.3cm}

$^2$PH-TH, CERN, CH--1211 Gen\`eve 23, Switzerland

\vspace{0.3cm}

$^3$Institut f\"ur Theoretische Physik,
Karlsruhe Institute of Technology, \\
D--76128 Karlsruhe, Germany

}
\end{center}

\vspace*{0.2cm}

\BC {\bf Abstract} \EC
The Higgs bosons of the MSSM can be produced from the decay of SUSY particles.
We review the evalulation of two decay modes in the MSSM with complex
parameters (cMSSM).
The first type is the decay of the heavy scalar top quark
to a lighter scalar quark and a Higgs boson.
The second type is the decay of the heavy chargino to a lighter
chargino/neutralino and a Higgs boson.
The evaluation is based on a full
one-loop calculation including hard QED and QCD radiation. 
We find sizable contributions to many partial decay widths and  
branching ratios. They are roughly of \order{10\%} of the tree-level 
results, but can go up to $30\%$ or higher. 
These contributions are important for the correct interpretation
of scalar top quark decays at a future linear $e^+e^-$
collider.

\def\thefootnote{\arabic{footnote}}
\setcounter{footnote}{0}

\newpage


\title{MSSM Higgs Bosons from Stop and Chargino Decays}
%
\author{S.~Heinemeyer$^1$, F.~v.d.~Pahlen$^1$, H.~Rzehak$^2$
and C.~Schappacher$^3$
\vspace{.3cm}\\
1- Instituto de F\'isica de Cantabria (CSIC),
E--39005 Santander, Spain
\vspace{.1cm}\\
2- PH-TH, CERN, CH--1211 Gen\`eve 23, Switzerland
\vspace{.1cm}\\
3- Institut f\"ur Theoretische Physik,
Karlsruhe Institute of Technology, \\
D--76128 Karlsruhe, Germany\\
}

\maketitle

\begin{abstract}
The Higgs bosons of the MSSM can be produced from the decay of SUSY particles.
We review the evalulation of two decay modes in the MSSM with complex
parameters (cMSSM).
The first type is the decay of the heavy scalar top quark
to a lighter scalar quark and a Higgs boson.
The second type is the decay of the heavy chargino to a lighter
chargino/neutralino and a Higgs boson.
The evaluation is based on a full
one-loop calculation including hard QED and QCD radiation. 
We find sizable contributions to many partial decay widths and  
branching ratios. They are roughly of \order{10\%} of the tree-level 
results, but can go up to $30\%$ or higher. 
These contributions are important for the correct interpretation
of scalar top quark decays at a future linear $e^+e^-$
collider.

\end{abstract}


\section{Introduction}

One of the most important tasks of current high-energy physics is the
search for physics 
effects beyond the Standard Model (SM), where the Minimal Supersymmetric
Standard Model (MSSM)~\cite{mssm} is one of the leading candidates. 
Supersymmetry (SUSY) predicts two scalar partners for all SM fermions as well
as fermionic partners to all SM bosons.
Another important task is the investigation and identification of the
mechanism of electroweak symmetry breaking. 
The most frequently investigated
models are the Higgs mechanism within the SM and within the MSSM.
Contrary to the case of the SM, in the MSSM 
two Higgs doublets are required.
This results in five physical Higgs bosons instead of the single Higgs
boson in the SM; three neutral Higgs bosons, $h_n$ ($n = 1,2,3$), 
and two charged Higgs bosons, $H^\pm$. 
In the MSSM with complex parameters (cMSSM) the three neutral Higgs
bosons mix~\cite{mhiggsCPXgen,mhiggsCPXRG1,Demir,mhiggsCPXFD1}, 
giving rise to the states $\He, \Hz, \Hd$.

An interesting production channel of Higgs bosons is 
the decay of the heavy scalar top  quark to the lighter scalar top 
(scalar bottom)  quark and a neutral (charged) Higgs boson.
Another SUSY particle that can produce a Higgs boson is a
chargino, which can decay to a lighter chargino (a lighter neutralino)
and a neutral (charged) Higgs boson.

The original heavier SUSY particles can be produced at the LHC, or if
kinematically allowed at an $e^+e^-$ collider.
At the ILC (or any other future $e^+e^-$ collider such as CLIC) 
a precision determination of the properties of the observed particles is
expected~\cite{teslatdr,ilc}. 
Thus, if kinematically accessible, Higgs production via scalar top quark
or chargino decays could offer important information about the Higgs
sector of the MSSM. 

In order to yield a sufficient accuracy, one-loop corrections to
the various SUSY decay modes have to be considered.
For the precise evaluation of the branching ratio at least all 
two-body decay modes have to be considered and evaluated at the one-loop
level. We review the results for the evaluation of these decay widths
(and branching ratios) obtained in the MSSM with complex parameters
(cMSSM)~\cite{Stop2decay,LHCxC}. 
We will review the numerical results for 
\begin{align}
\label{ststphi}
&\Ga(\Stopz \to \Stope h_n) \qquad (n = 1,2,3)~, \\
\label{stsbH}
&\Ga(\Stopz \to \Sbote H^+) ~, \\
\label{C2NH}
&\Ga(\DecayCNH{2}{j}) \qquad (j = 1,2,3)~, \\
\label{C1NH}
&\Ga(\DecayCNH{1}{1}) ~, 
\end{align}
where $\neu{k}$ denotes the neutralinos, $\cha{j}$
the charginos. 
The total decay width is defined as the sum of {\em all} the partial 
two-body decay widths, which have {\em all} be evaluated at the one-loop level.

We also concentrate on the decays of $\Stopz, \cham{2,1}$ and do not
investigate $\aStopz, \chap{2,1}$ decays. 
In the presence of complex phases this
would lead to somewhat different results. 
Detailed references to existing calculations of these decay widths, 
branching ratios, as well about the extraction of complex phases can be
found in \citeres{Stop2decay,LHCxC}.  
Our results will be  implemented into the Fortran code 
{\tt FeynHiggs}~\cite{feynhiggs,mhiggslong,mhiggsAEC,mhcMSSMlong}.


\section{The complex MSSM and its renormalization}
\label{sec:renorm}

All the relevant two-body decay channels are evalulated at the
one-loop level, including hard QED and QCD radiation. This requires the
simultaneous renormalization of several sectors of the cMSSM, including
the colored sector with top and bottom quarks and their scalar partners
as well as the gluon and the gluino, the Higgs and gauge boson sector with 
all the Higgs bosons as well as the $Z$ and the $W$~boson and the
chargino/neutralino sector. Details about our notation and especially
about the renormalization of the
cMSSM can be found in \citeres{SbotRen,Stop2decay,LHCxC,Gluinodecay}. 

An important role play contributions of 
self-energy type of external (on-shell) particles. While the real part of 
such a loop
does not contribute to the decay width due to the on-shell
renormalization, the imaginary part, in product with an imaginary part
of a complex coupling (such as $\At$ or $M_1$) can give a real
contribution to the decay width. These contributions (in the following called
``absorptive contributions'')
have been taken into account in the analytical and numerical
evaluation.
The impact of those contributions will be discussed in 
\refses{sec:numevalstop}, \ref{sec:numevalchar}.

The Feynman diagrams and corresponding amplitudes contributing to the
various decays have been obtained with \fa~\cite{feynarts}. 
The model file, including the MSSM counterterms, 
is largely based on \citere{dissTF}, however adjusted to
match exactly the renormalization prescription described in
\citere{SbotRen,Stop2decay,LHCxC,Gluinodecay}. 
The further evaluation has been performed with \fc~\cite{formcalc}. 
As regularization scheme for the UV-divergences we
have used constrained differential renormalization~\cite{cdr}, 
which has been shown to be equivalent to 
dimensional reduction~\cite{dred} at the \onel\ level~\cite{formcalc}. 
Thus the employed regularization scheme preserves SUSY~\cite{dredDS,dredDS2}. 
All UV-divergences cancel in the final result.
(Also all IR-divergences cancel in the one-loop result as required.)


{
\section{Numerical results for scalar top decays}
\label{sec:numevalstop}

\newcommand{\SE}{S1}
\newcommand{\SZ}{S2}

The numerical examples are shown in two numerical scenarios, \SE\ and
\SZ, where the parameters are given in \refta{tab:para}. 
The results shown in this section consist of 
``tree'', which denotes the tree-level value and of ``full'', which is
the partial decay width including {\em all} one-loop 
corrections. We only show the results for the decay widths, since the size
of the loop corrections to the branching ratios are more parameter
dependent. 

\begin{table}[htb!]
\renewcommand{\arraystretch}{1.2}
\BC
\begin{tabular}{|c||c|c|c|c|c|c|c|c|c|c|c|}
\hline
Scen.\ & $\tb$ & $\MHp$ & $\mstz$ & $\mste$  & $\msbz$ 
& $\mu$ & $\At$ & $\Ab$ & $M_1$ & $M_2$ & $M_3$ 
\\ \hline\hline
\SE & 20 & 150 & 650 & $0.4\, \mstz$ & $0.7\, \mstz$ & 
200 &  800 &  400 & 200 & 300 & 350 
\\ \hline
\SZ & 20 & 180 & 1200 & $0.6\, \mstz$ & $0.8\, \mstz$ &
300 & 1800 & 1600 & 150 & 200 & 400  
\\ \hline
\end{tabular}
\caption{MSSM parameters for the initial numerical investigation; all
  masses are in GeV. 
}
\label{tab:para}
\EC
\renewcommand{\arraystretch}{1.0}
\end{table}

The production of $\Stopz$ at the ILC(1000), i.e.\ with 
$\sqrt{s} = 1000 \gev$, via $e^+e^- \to \aStope\Stopz$ will be possible,
with all the decay modes (\ref{ststphi}), (\ref{stsbH})
being open. The clean environment of the ILC would permit a detailed
study of the scalar top decays.
For the parameters in \refta{tab:para} we find 
$\si(e^+e^- \to \aStope\Stopz) \approx 1.4~{\rm fb}$, 
i.e.\ an integrated
luminosity of $\sim 1\, \iab$ would yield about~1400~$\Stopz$. 
The ILC environment would result in an accuracy of
the relative branching ratio close to the statistical
uncertainty: a BR of 30\% could be determined to $\sim 6\%$ for the
$\mstz$ values in \refta{tab:para}. 
Depending on the combination of allowed decay
channels a determination of the branching ratios at the few per-cent
level might be achievable in the high-luminosity running of the ILC(1000).

We show the results for the various decay widths as a function of $\phiat$. 
The other parameters are chosen according to \refta{tab:para}. 
Thus, within \SE\ we have $\mste + \mstz = 910 \gev$, i.e.\ 
the production channel $e^+e^- \to \aStope\Stopz$ 
is open at the ILC(1000). 
Consequently, the accuracy of the prediction of the various partial decay 
widths and branching ratios should be at the same level (or better) as 
the anticipated ILC precision.

\begin{figure}[htb!]
\begin{center}
\includegraphics[width=0.45\textwidth,height=4cm]{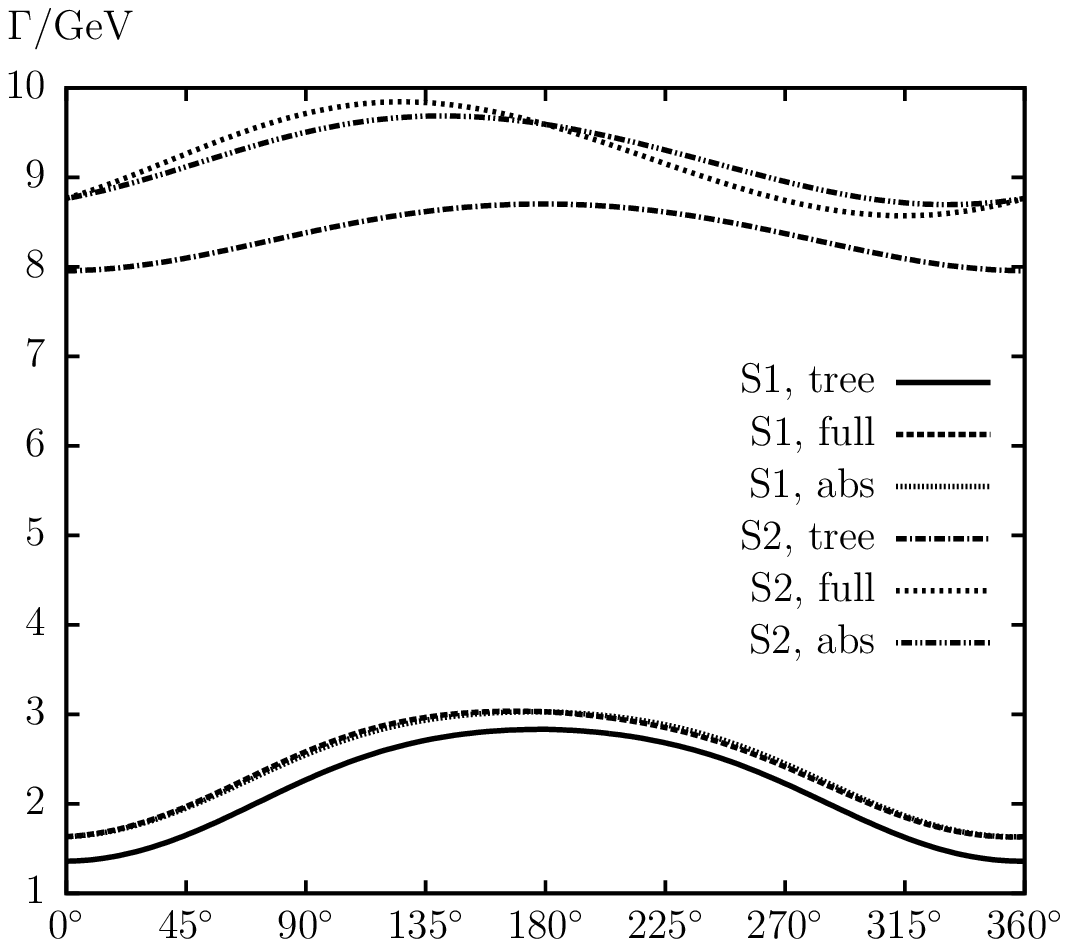}
\hspace{-4mm}
\includegraphics[width=0.45\textwidth,height=4cm]{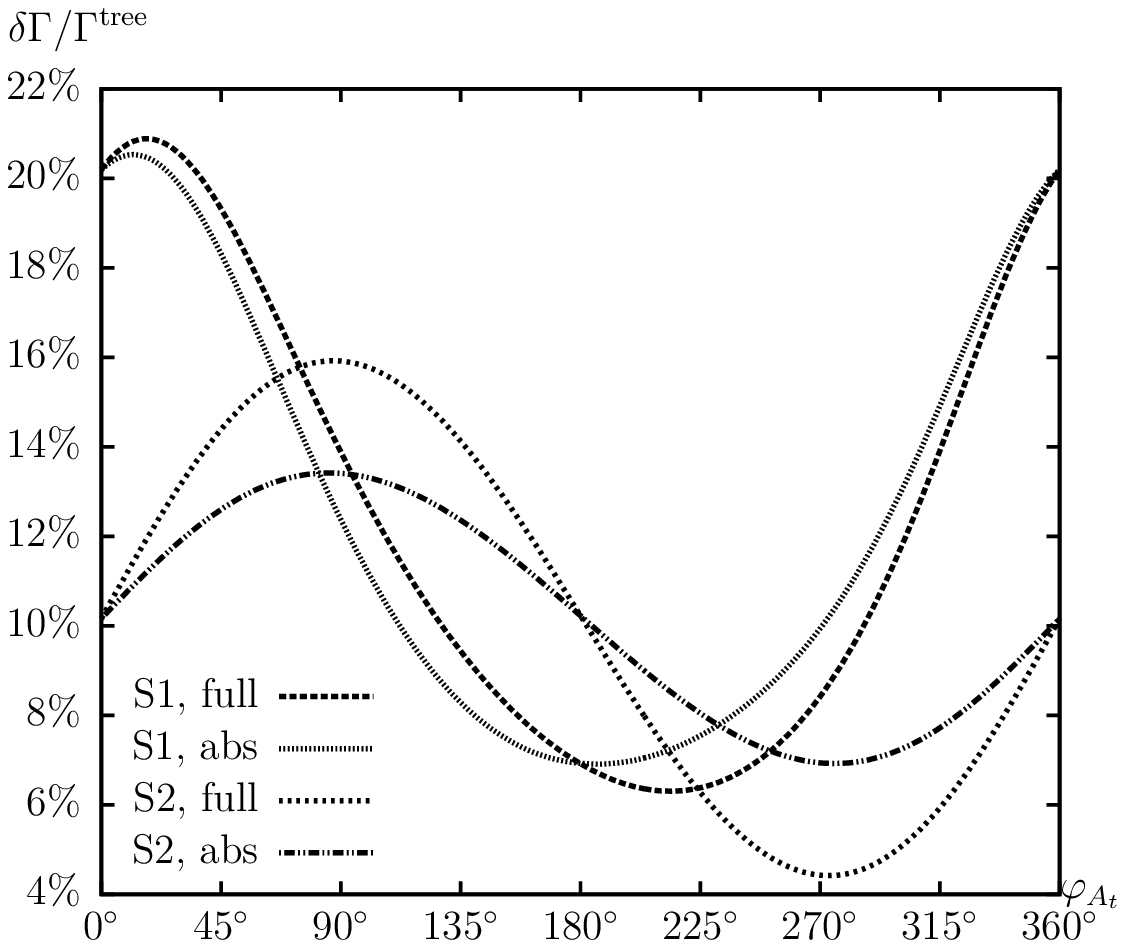}\\[.2em]
\includegraphics[width=0.45\textwidth,height=4cm]{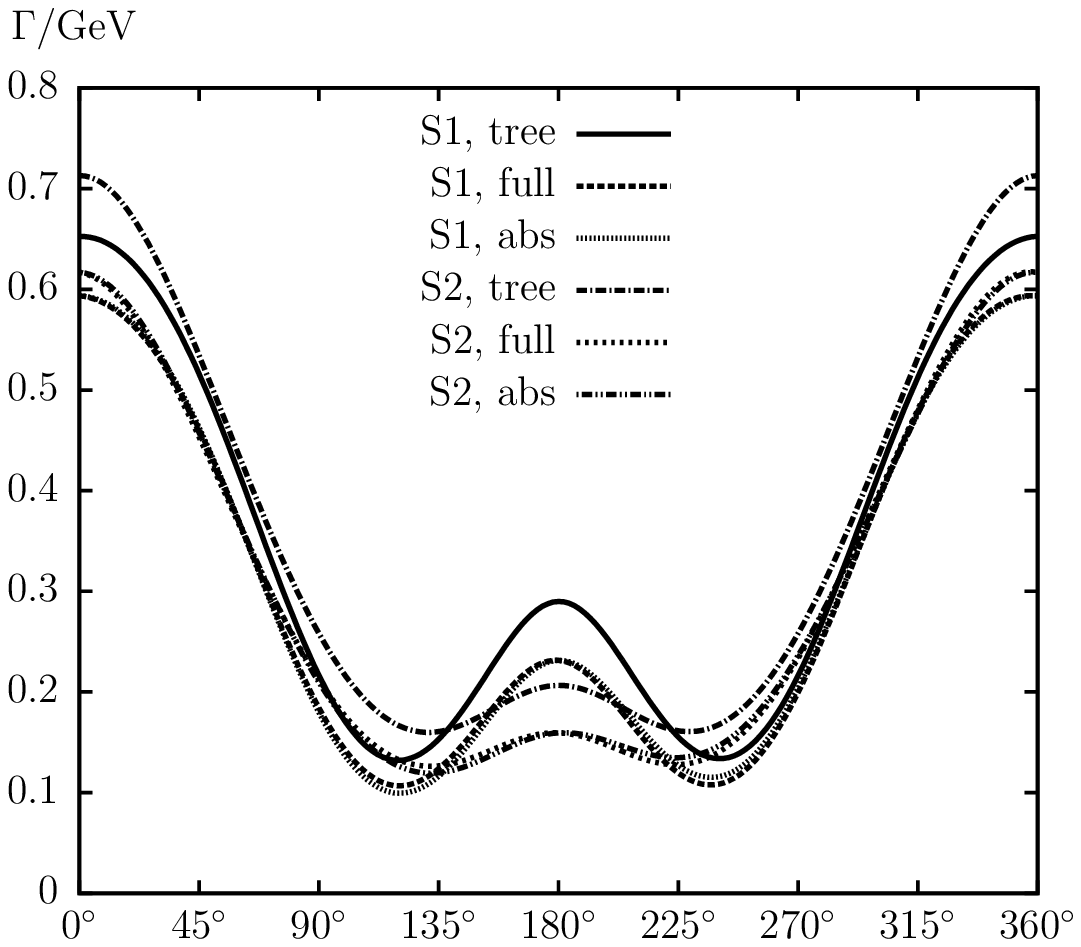}
\hspace{-4mm}
\includegraphics[width=0.45\textwidth,height=4cm]{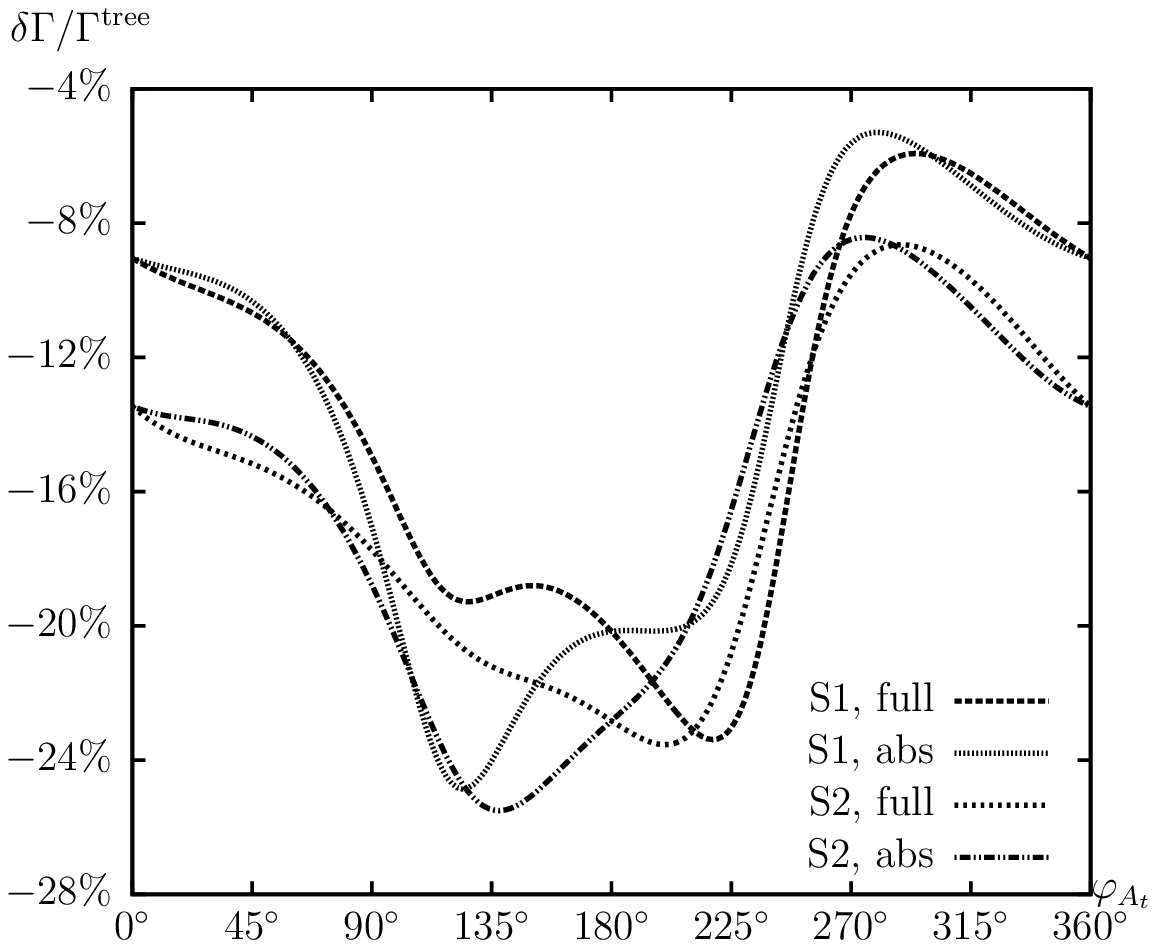}\\[.2em]
\includegraphics[width=0.45\textwidth,height=4cm]{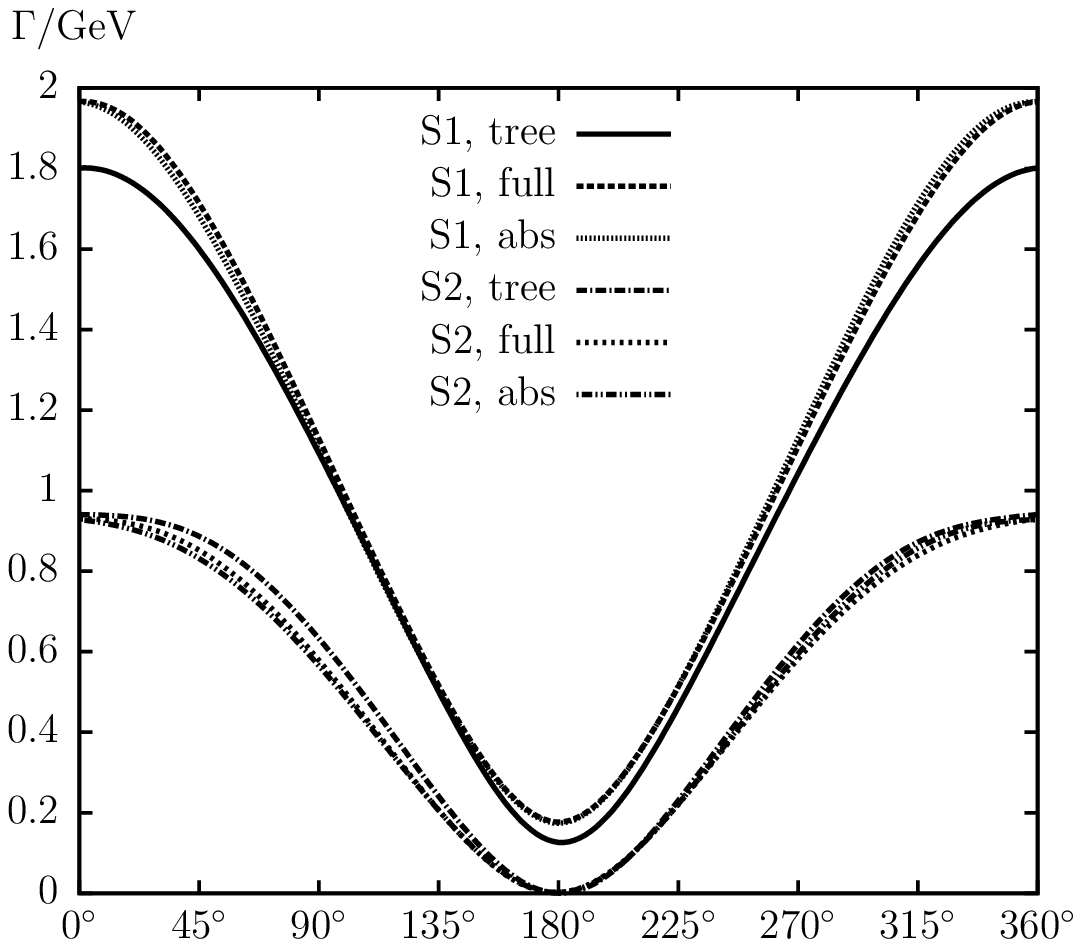}
\hspace{-4mm}
\includegraphics[width=0.45\textwidth,height=4cm]{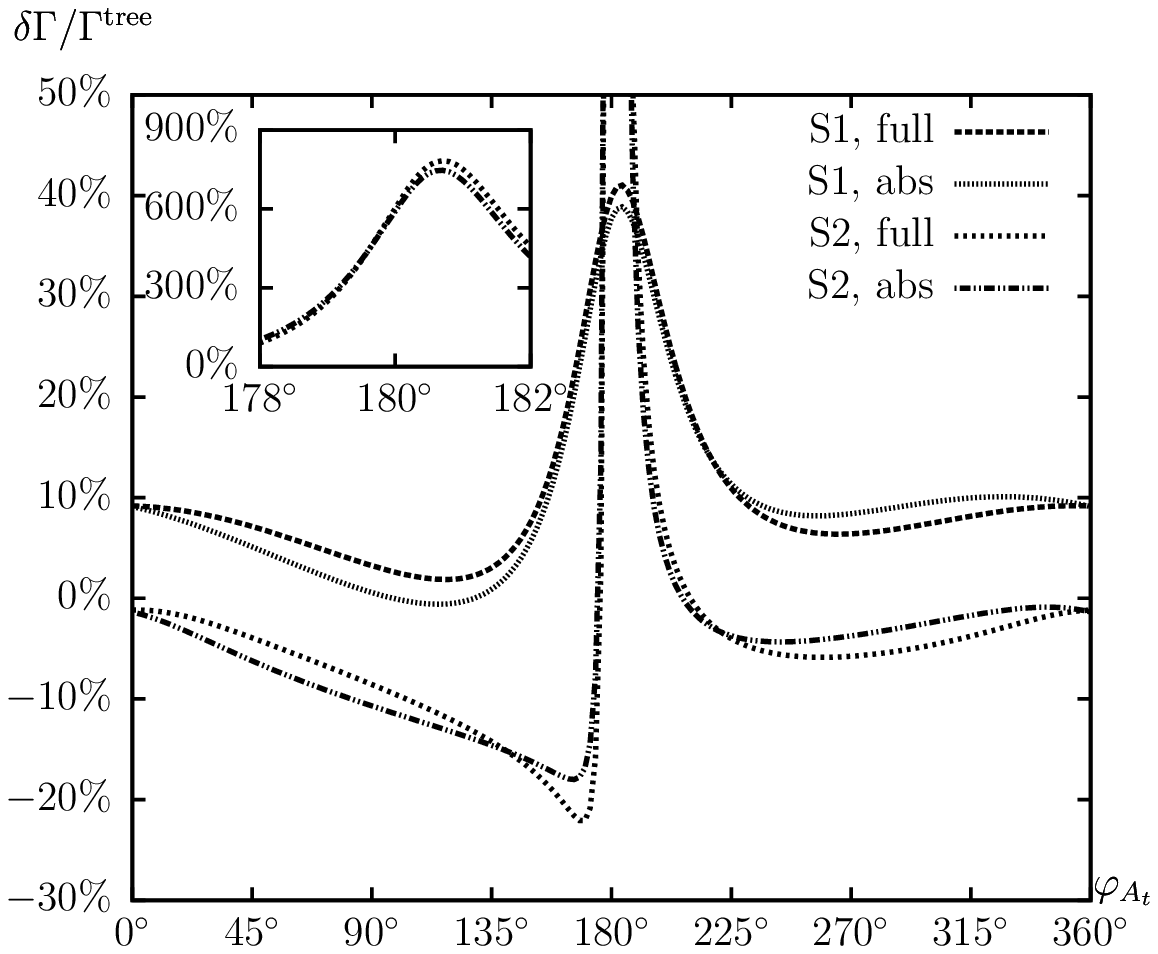}\\[.2em]
\includegraphics[width=0.45\textwidth,height=4cm]{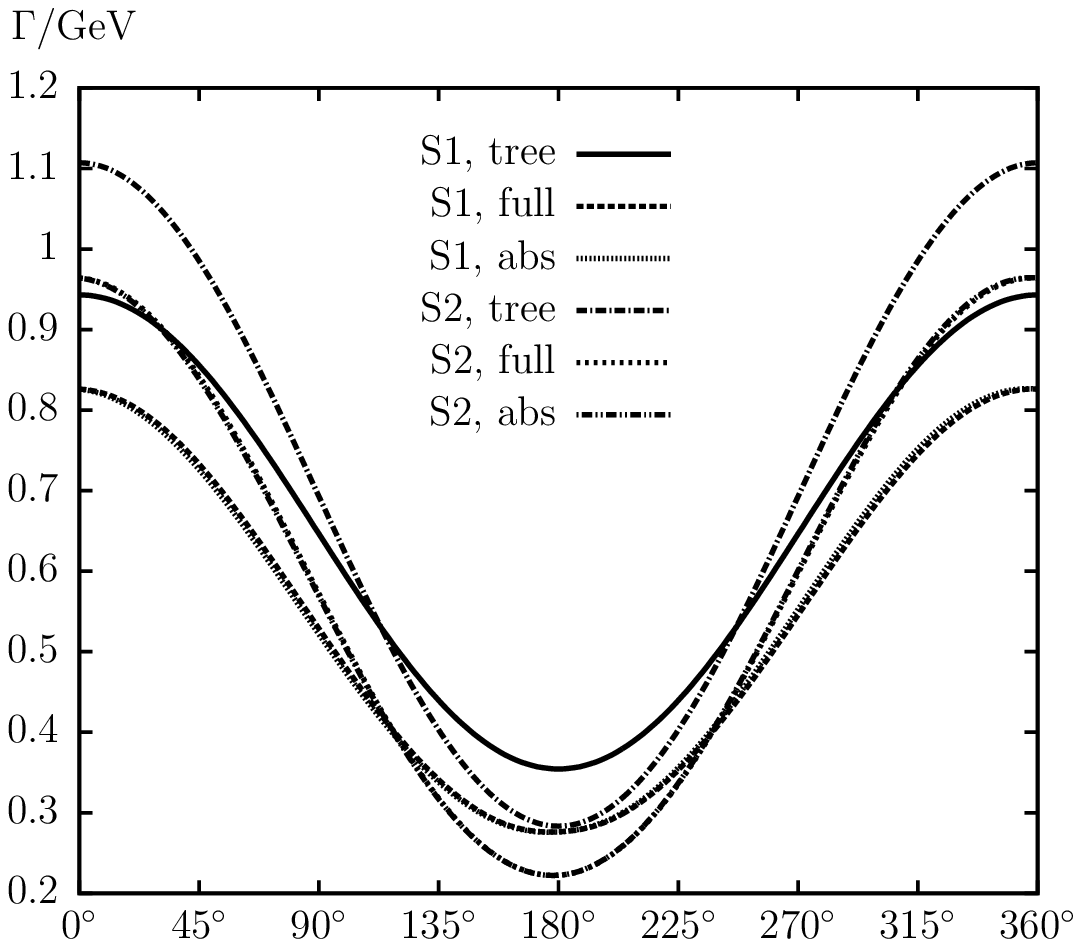}
\hspace{-4mm}
\includegraphics[width=0.45\textwidth,height=4cm]{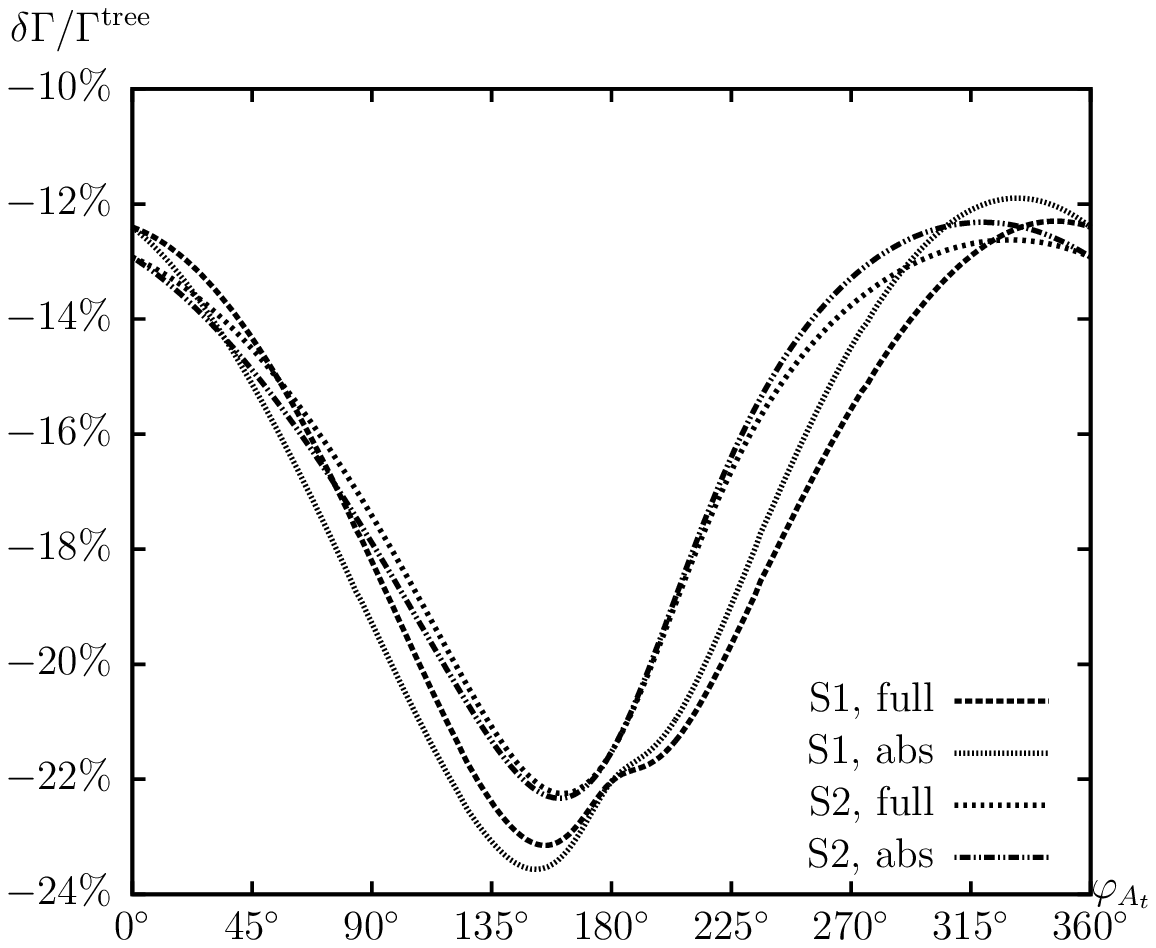}
\caption{
  Tree-level (``tree'') and full one-loop (``full'') corrected 
  partial decay widths are shown with $\phiat$ varied. 
  Also shown are the full one-loop
  corrected partial decay  
  widths including absorptive contributions (``abs''). 
  First row: $\Ga(\decayh)$, second row: $\Ga(\decayH)$, third row:
  $\Ga(\decayA)$, fourth row: $\Ga(\decaySbeH)$.
}
\label{fig:PhiAt.st2st1h}
\end{center}
\end{figure}

In \reffi{fig:PhiAt.st2st1h} we show $\Ga(\decayh)$ (first), 
$\Ga(\decayH)$ (second), $\Ga(\decayA)$ (third) and $\Ga(\decaySbeH)$ 
(fourth row) as a function of $\phiat$ for the parameters in
\refta{tab:para}, where the left (right) column displays the (relative
one-loop correction to the) decay width. While $\Ga(\decayh)$ in \SZ\
is of \order{9 \gev}, the other decay widths shown are of \order{1 \gev}.
The variation with $\phiat$ can be seen to be very large, of 
\order{50\%}. The size of the one-loop corrections, as shown in the
right column, is also sizable, of \order{\pm 20\%} and exhibit a strong
variation with $\phiat$. The effects of the ``absorptive contributions''
are clearly visible, especially for $\decayh$.
Consequently, the full one-loop corrections 
must be taken into account in a reliable complex phase
determination from scalar top decays.

}


{

\newcommand{\SN}{${\cal S}$}
\newcommand{\SE}{\ensuremath{{\cal S}_>}}
\newcommand{\SZ}{\ensuremath{{\cal S}_<}}

\section{Numerical results for chargino decays}
\label{sec:numevalchar}

The numerical examples are evaluated using the parameters given in 
\refta{tab:para-cn}. We assume the scalar quarks to be sufficiently heavy such
that they do 
not contribute to the total decay widths of the charginos.
We invert the expressions of the chargino masses 
in order to express the parameters
$\mu$ and $\MTwo$ (which are taken to be real) as a
function of $\mcha{1}$ and $\mcha{2}$.
This leaves two choices for the hierarchy of $\mu$ and $\MTwo$:
\begin{align}
\label{eq.SE}
\SE &: \mu > \MTwo \quad (\cha{2} \mbox{~more higgsino-like})~, \\
\label{eq.SZ}
\SZ &: \mu < \MTwo \quad (\cha{2} \mbox{~more gaugino-like})~.
\end{align}
The absolute value of $\MOne$ is fixed via the GUT
relation (with $|\MTwo| \equiv \MTwo$)
\begin{align}
|\MOne| &= \frac{5}{3} \tan^2 \thw \MTwo \approx \edz \MTwo~,
\label{M1M2}
\end{align}
leaving $\phiMe$ as a free parameter.

\begin{table}[t!]
\renewcommand{\arraystretch}{1.2}
\BC
\begin{tabular}{|c||c|c|c|c|c|c|c|c|}
\hline
Scen.\ & $\tb$ & $\MHp$ & $\mcha{2}$ & $\mcha{1}$ 
       & $\MslL$ & $\MslR$ & $\Al$ 
\\ \hline\hline
\SN & 20 & 160 & 600 & 350 & 300 & 310 & 400 
\\ \hline
\end{tabular}
\caption{MSSM parameters for the numerical investigation of chargino
  decays; all masses are in GeV. 
}
\label{tab:para-cn}
\EC
\renewcommand{\arraystretch}{1.0}
\end{table}

The values of $\mcha{1,2}$ allow $\cha{1}\champ{2}$ or $\chap{1}\cham{1}$
production at the ILC(1000) via
$e^+e^- \to \cha{1}\champ{1,2}$,
with all the subsequent decay modes to a neutralino and a charged Higgs
boson, see \refeqs{C2NH}, (\ref{C1NH}).
As for the scalar top decays the clean environment of the ILC would
permit a detailed study of the chargino decays.
For the values in \refta{tab:para-cn} and unpolarized beams
we find, for $\SE$ ($\SZ$),
$\si(e^+e^- \to \cha{1}\champ{2}) \approx 4\, (12)~{\rm fb}$, and
$\si(e^+e^- \to \chap{1}\cham{1}) \approx 55\, (80)~{\rm fb}$. 
Choosing appropriate polarized beams these cross sections can be
enhanced by a factor of approximately $2$ to $3$.
An integrated luminosity of $\sim 1\, \iab$ would yield about 
$4-12 \times 10^3$ $\cha{1}\champ{2}$ events and about
$55 - 80 \times 10^3$ $\chap{1}\cham{1}$ events, with appropriate
enhancements in the case of polarized beams.
The ILC environment would result in an accuracy of
the relative branching ratio close to the statistical
uncertainty, see the previous section.
Depending on the combination of allowed decay channels a determination of 
the branching ratios at the per-cent level might be achievable in the 
high-luminosity running of the ILC(1000).

The results shown in this section consist of 
``tree'', which denotes the tree-level value and of ``full'', which is
the partial decay width including {\em all} one-loop 
corrections. 
Also shown are the full one-loop corrected decay widths omitting
the absorptive contributions (``full R'').
We only show the results for the decay widths, since the size
of the loop corrections to the branching ratios are more parameter
dependent.

\begin{figure}[htb!]
\begin{center}
\includegraphics[width=0.45\textwidth,height=4cm]{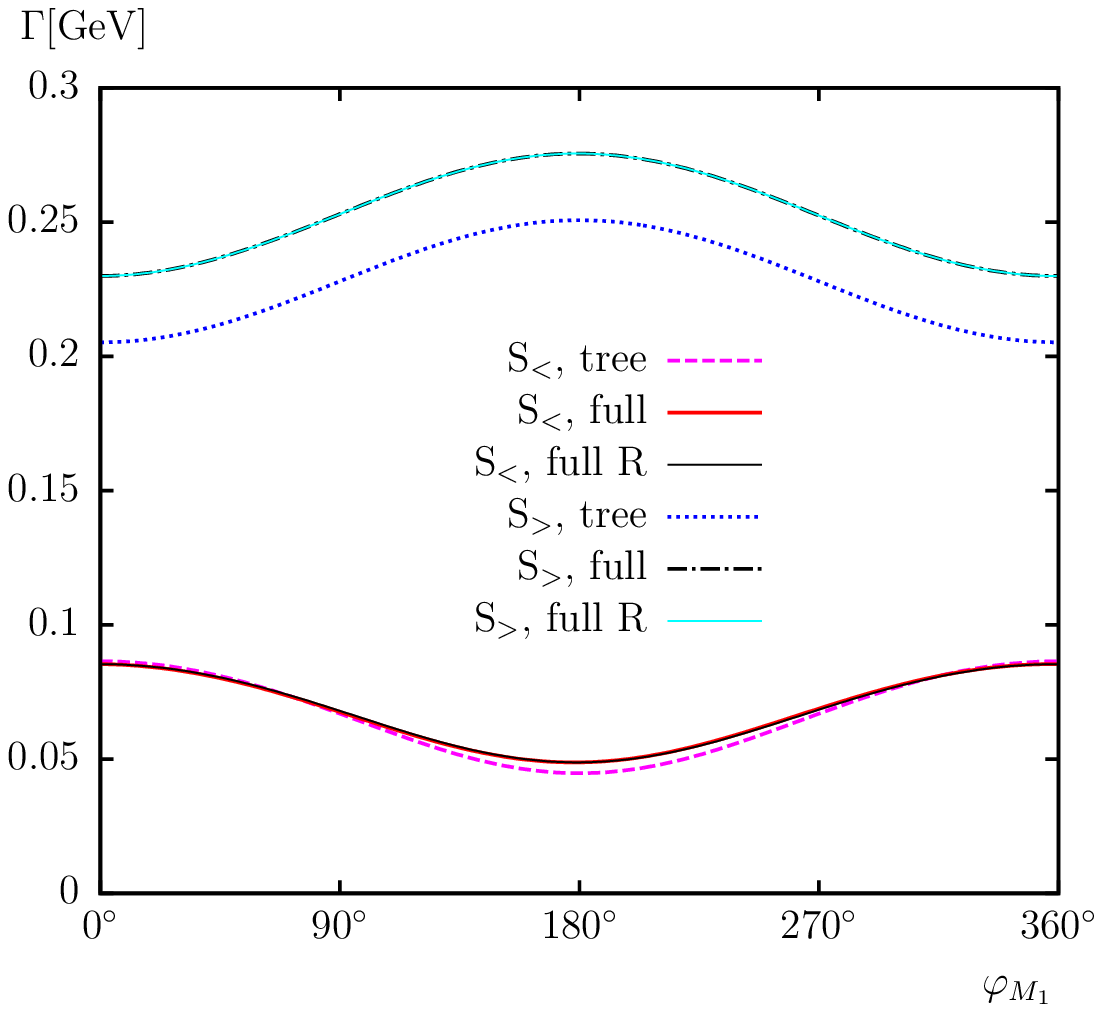}
\hspace{-4mm}
\includegraphics[width=0.45\textwidth,height=4cm]{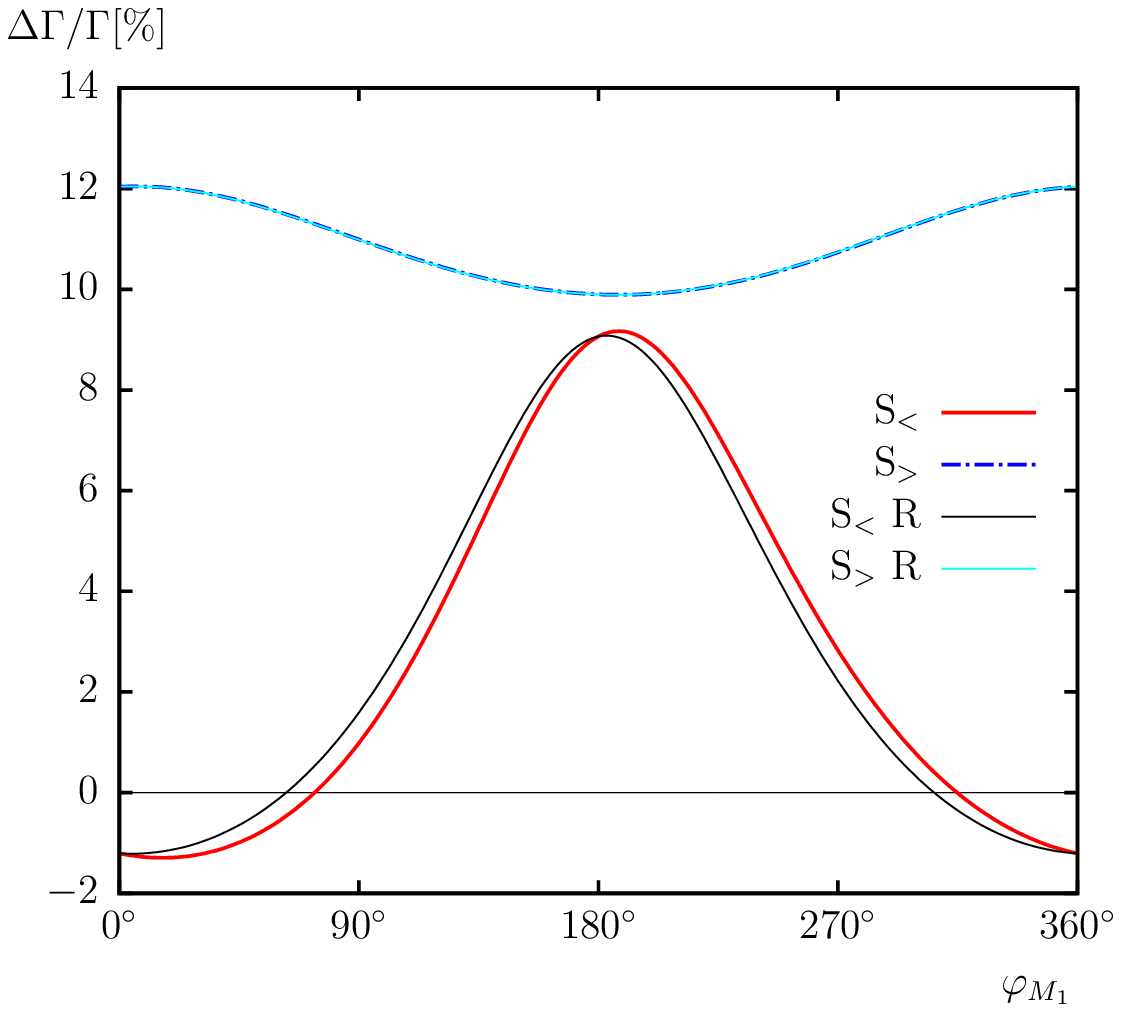}\\[.2em]
\includegraphics[width=0.45\textwidth,height=4cm]{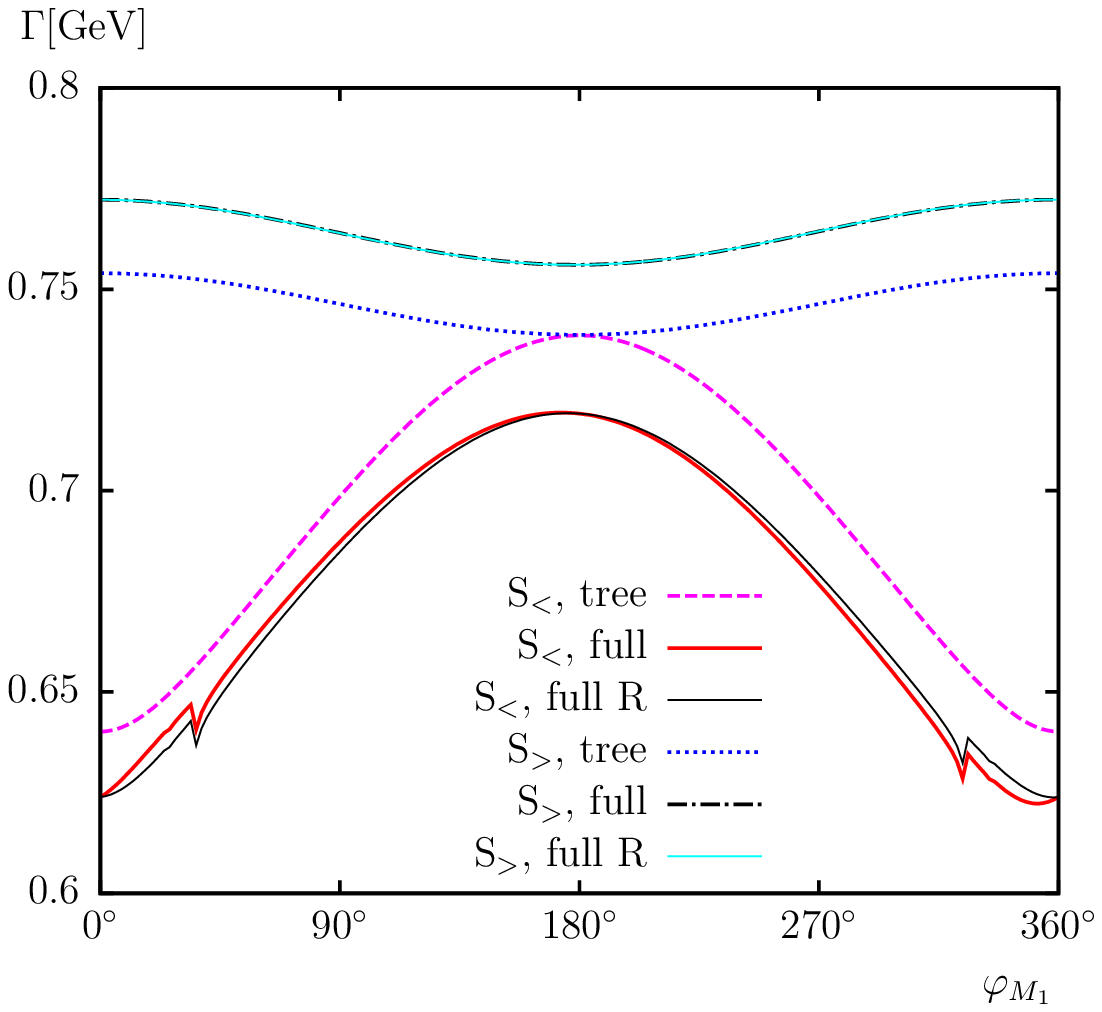}
\hspace{-4mm}
\includegraphics[width=0.45\textwidth,height=4cm]{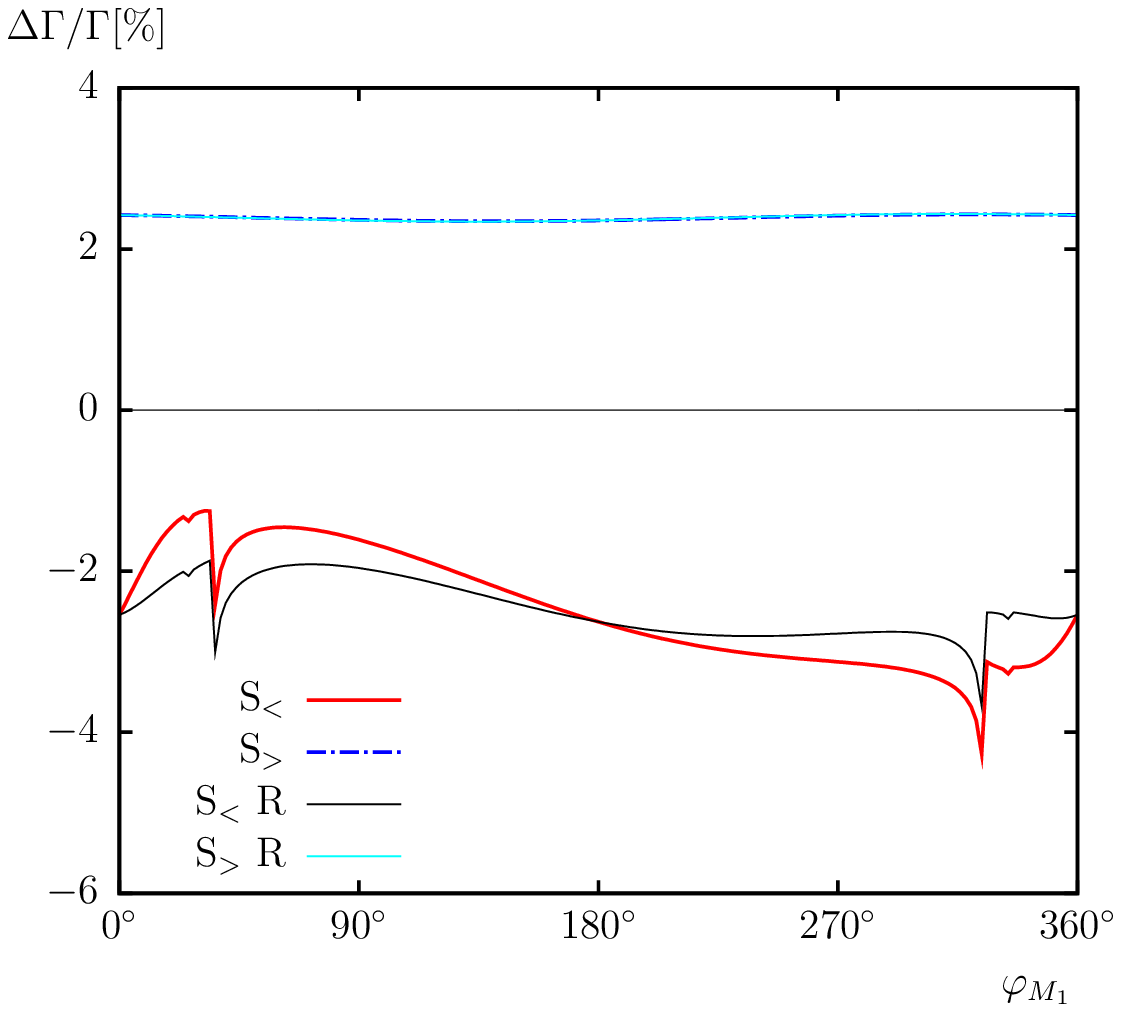}\\[.2em]
\includegraphics[width=0.45\textwidth,height=4cm]{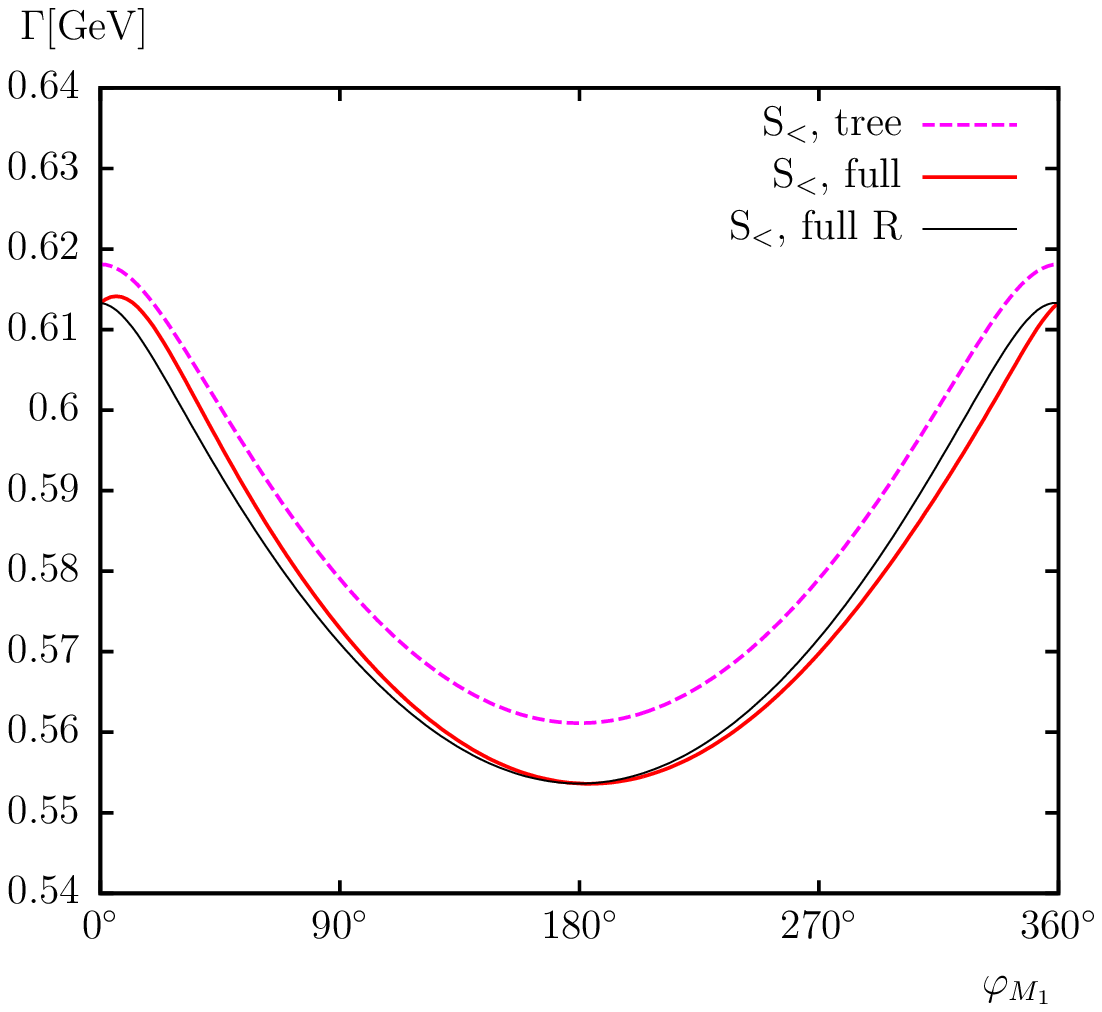}
\hspace{-4mm}
\includegraphics[width=0.45\textwidth,height=4cm]{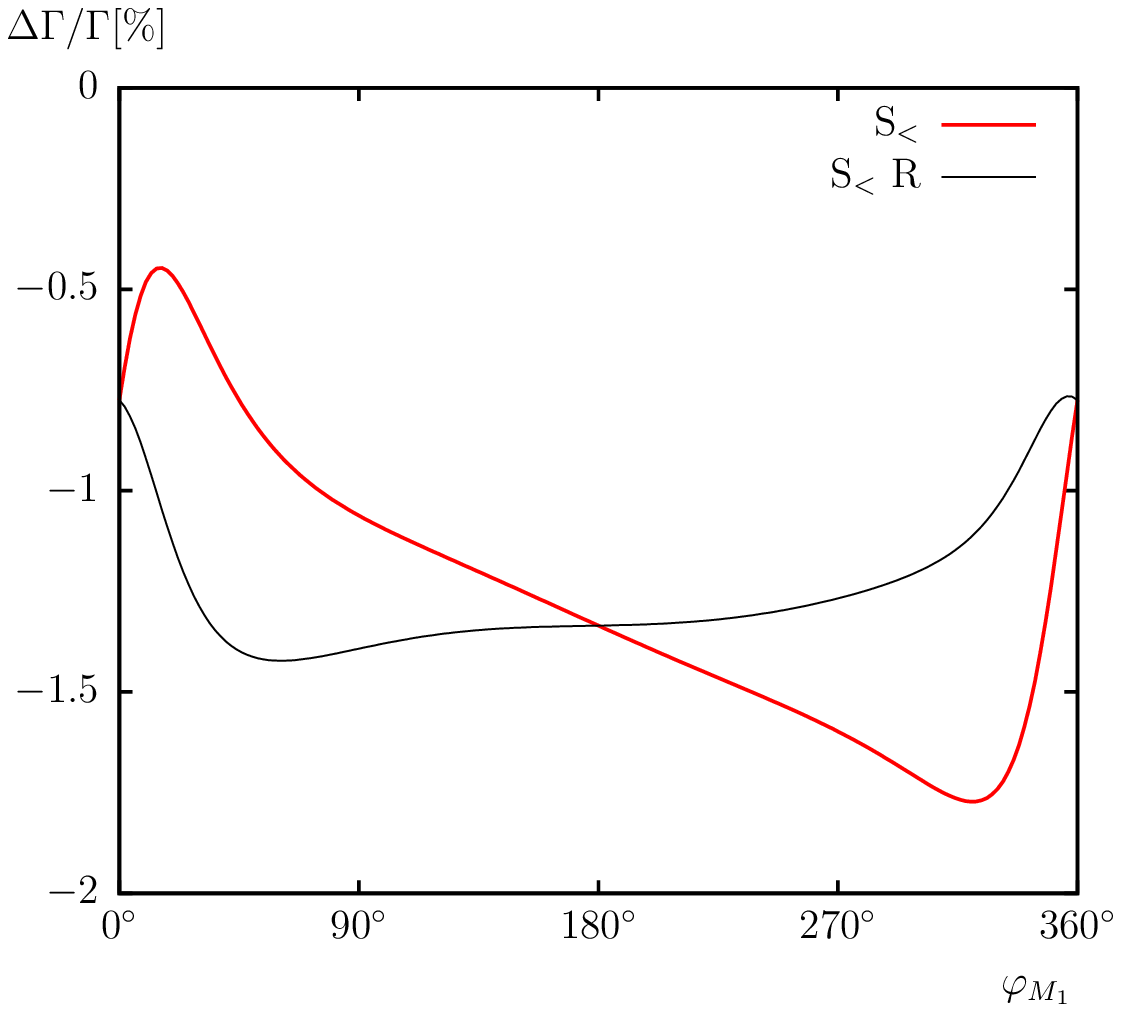}\\[.2em]
\includegraphics[width=0.45\textwidth,height=4cm]{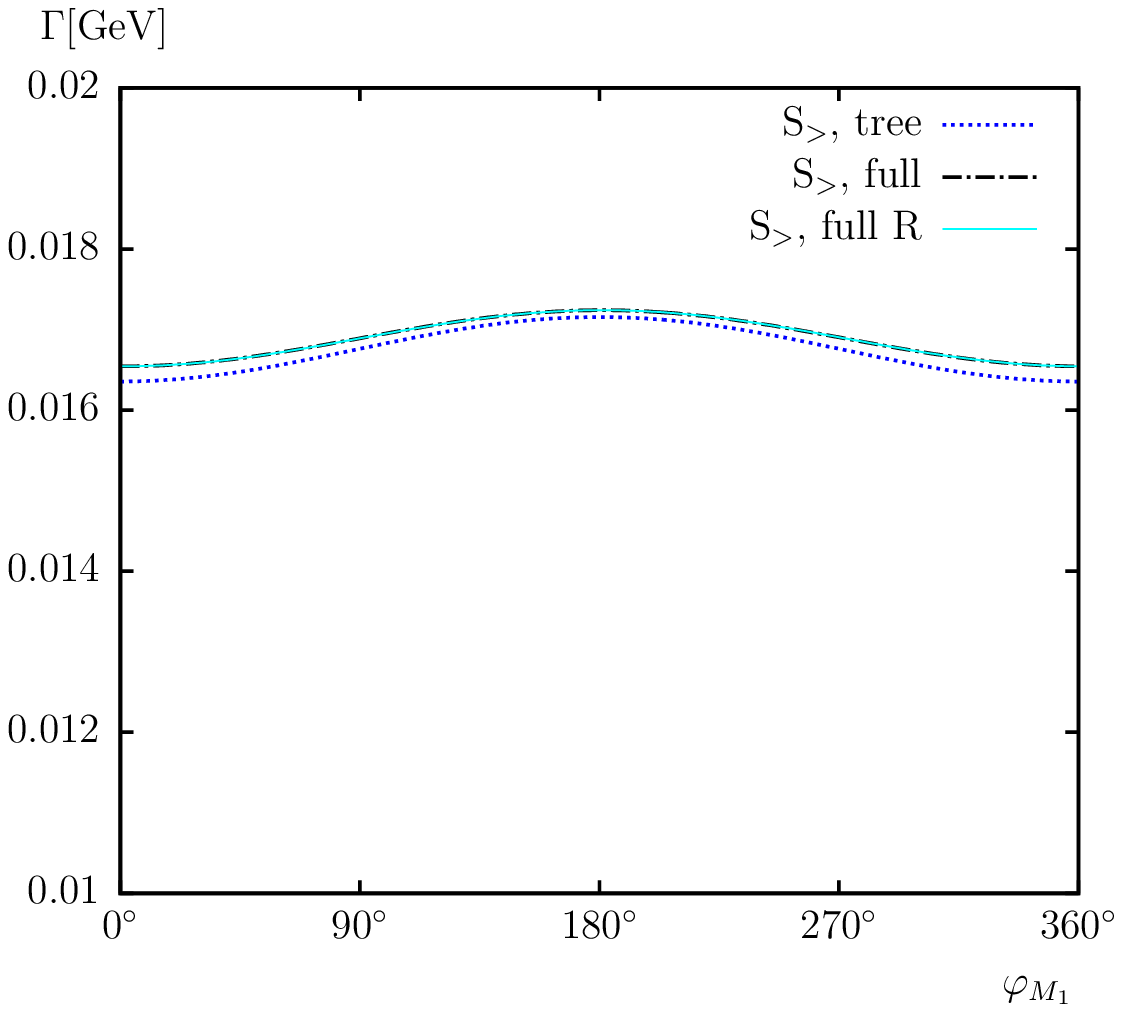}
\hspace{-4mm}
\includegraphics[width=0.45\textwidth,height=4cm]{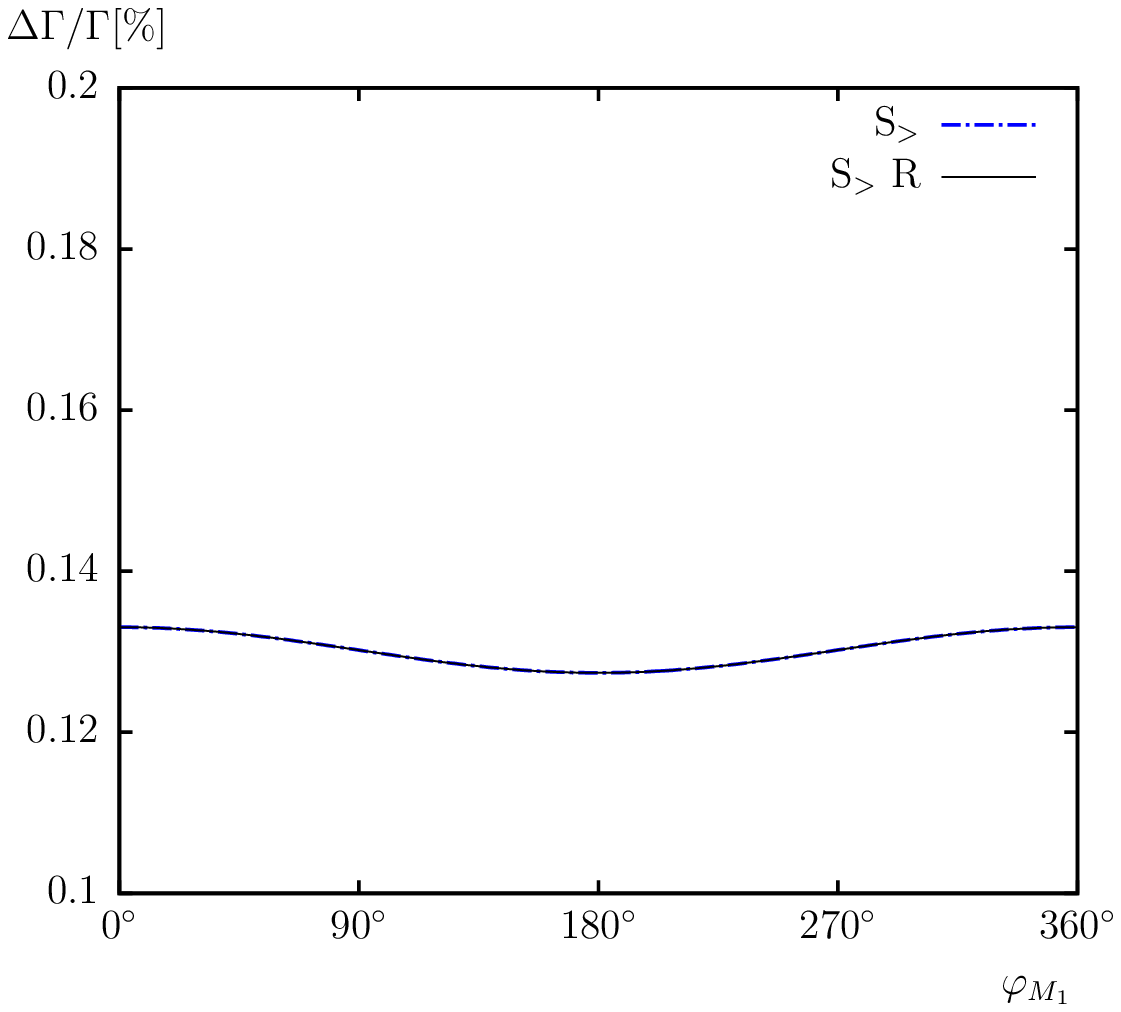}
\caption{
  Tree-level (``tree'') and full one-loop (``full'') corrected 
  decay widths are shown with $\phi_{\MOne}$ varied.
  Also shown are the full one-loop corrected decay widths omitting
  the absorptive contributions (``full R'').
  First row: $\Ga(\DecayCNH{2}{1})$,
  second row: $\Ga(\DecayCNH{2}{2})$,
  third row: $\Ga(\DecayCNH{2}{3})$,
  fourth row: $\Ga(\DecayCNH{1}{1})$.
}
\label{fig:PhiM1.chaneuhp}
\end{center}
\end{figure}

In \reffi{fig:PhiM1.chaneuhp} we show $\Ga(\DecayCNH{2}{1})$ (first), 
$\Ga(\DecayCNH{2}{2})$ (second), $\Ga(\DecayCNH{2}{3})$ (third) 
and $\Ga(\DecayCNH{1}{1})$ 
(fourth row) as a function of $\phiMe$ for the parameters in
\refta{tab:para-cn}, where the left (right) column displays the (relative
one-loop correction to the) decay width. 
The decay widhts are of \order{0.1 \gev} in the case of
$\DecayCNH{2}{1}$, about five times larger for $\DecayCNH{2}{2,3}$ and a
factor of ten smaller for the light chargino decay. 
For the heavy chargino decay a strong variation with $\phiMe$ can be
observed. The size of the one-loop corrections, as shown in the
right column, is also sizable in the case of the heavy chargino, 
between $-4\%$ and $+12\%$ and show a non-negligible dependence on
$\phiMe$. Again the effects of the ``absorptive contributions'' are
clearly visible. Also these loop corrections should be
taken into account in a reliable complex phase determination in the
chargino/neutralino sector.

}


\section*{Acknowledgments}

The work of S.H.\ was partially supported by CICYT (grant FPA
2007--66387 and FPA 2010--22163-C02-01).
F.v.d.P.\ was supported by 
the Spanish MICINN's Consolider-Ingenio 2010 Programme under grant
MultiDark CSD2009-00064.



\begin{footnotesize}


\end{footnotesize}

\end{document}